\documentclass[draftcls,onecolumn]{IEEEtran}
\usepackage[utf8]{inputenc}
\usepackage{mathtools,amssymb}
\usepackage{graphicx}
\usepackage{mathrsfs}
\usepackage[dvipsnames]{xcolor}
\colorlet{bvb}{Red}
\colorlet{wz}{blue}
\colorlet{bvb2}{orange}

\DeclareMathAlphabet\mathbfcal{OMS}{cmsy}{b}{n}

\DeclareRobustCommand*{\IEEEauthorrefmark}[1]{%
  \raisebox{0pt}[0pt][0pt]{\textsuperscript{\footnotesize #1}}%
}
\begin{document}

\title{Machine Learning for Model Order Selection in MIMO 
OFDM Systems}

\author{
\IEEEauthorblockN{\textit{Brenda Vilas Boas\IEEEauthorrefmark{1}$^,$\IEEEauthorrefmark{2}, Wolfgang Zirwas\IEEEauthorrefmark{1}},
\textit{Martin Haardt\IEEEauthorrefmark{2}}}\\\ 
\IEEEauthorblockA{\IEEEauthorrefmark{1}Nokia, Germany \\\ \IEEEauthorrefmark{2}Ilmenau University of Technology, Germany }}


\maketitle

\begin{abstract}
A variety of wireless channel estimation methods, e.g., MUSIC and ESPRIT, rely on prior knowledge of the model order. Therefore, it is important to correctly estimate the number of multipath components (MPCs) which compose such channels. However, environments with many scatterers
may generate MPCs which are closely spaced. This clustering of MPCs in addition to noise makes the model order selection task difficult in practice to currently known algorithms. In this paper, we exploit the multidimensional characteristics of MIMO orthogonal frequency division multiplexing (OFDM) systems and propose a machine learning (ML) method capable of 
determining the number of MPCs with a higher accuracy than state of the art methods
in almost coherent scenarios. Moreover, our results show that our proposed ML method has an enhanced reliability.

\end{abstract}

\begin{IEEEkeywords}
Source number detection, MIMO, deep learning, pattern recognition.
\end{IEEEkeywords}

\section{Introduction}

Model order estimation has been an active research field since four decades, especially due to its importance for parameter estimation algorithms and difficulty in coherent scenarios. Coherent scenarios happen when there is a high correlation between the signal sources. Many algorithms for estimating the model order 
rely on information theoretic criteria (ITC), such as the Akaike information criterion (AIC)~\cite{74akaikeAIC}, and the minimum description length (MDL)~\cite{78rissanenMDL}. However, those methods often fail when the number of snapshots is limited. 

The exponential fitting test (EFT)~\cite{06quinlanEFT} was proposed to overcome this problem by looking into the gap between the signal and the noise eigenvalues. The EFT finds a theoretical profile of the noise eigenvalues and recursively tests whether there is a mismatch on the observed eigenvalue and the theoretical one. A mismatch greater than a threshold indicates the presence of a source. Moreover, the EFT as well as AIC and MDL were extended to their higher dimensional version in~\cite{07costaMEFT}.     
Recently, the linear regression of global eigenvalues (LaRGE) method~\cite{21KorobkovLaRGE} has proposed to use the higher-order singular values computed by the higher-order singular value decomposition (HOSVD)~\cite{08HaardtTensor} to construct the global eigenvalues~\cite{07costaMEFT}. Those are further used to fit a linear curve to the noise eigenvalues, a relative prediction error above a defined threshold defines the model order. LaRGE is similar to EFT and
modified EFT (M-EFT)~\cite{07costaMEFT}; however, it does not require to compute the probability of false alarm.  

Lately, the increasing availability of computational capabilities has drawn attention to machine learning (ML), and, especially, deep neural networks (NNs) techniques. The model order selection problem has been tackled by ML methods in~\cite{20BarthelmeMOS,20YangMOS}. The work in~\cite{20BarthelmeMOS} proposes a NN for model order selection as a multiclass classification problem. The NN in~\cite{20BarthelmeMOS} takes the channel covariance matrix as input and is trained to output the model order. However, the proposed NN architecture has a large number of training parameters. In~\cite{20YangMOS}, the source number detection is modeled as a regression (ERNet) as well as multiclass classification (ECNet) task. Their 
NNs are trained to output the model order from the knowledge of the eigenvalues of the channel covariance matrix. Nonetheless, \cite{20YangMOS} does not take advantage of the multidimensional characteristics of MIMO systems, as the method just consider the eigenvalues in the spatial domain.   

Motivated by the multidimensional characteristics of MIMO orthogonal frequency division multiplexing (OFDM) systems,
we propose a new ML method which tackles the model order estimation problem as a multi-label classification task. Furthermore, we propose to use higher-order singular values as input to our ML to enhance the classification performance 
for radio channels with closely spaced multipath components (MPCs). 
Moreover, our proposed NN architecture can be easily adapted for channel models with higher dimensions, and more MPCs. As baseline methods, we compare to  LaRGE~\cite{21KorobkovLaRGE} as a non-ML method, and to the ECNet~\cite{20YangMOS} as a ML-method. Our results shows that our ML method has an increased accuracy. 

In this paper, 
Section~\ref{sec:scenario} presents our system overview and the wireless channel model, Section~\ref{sec:method} introduces our proposed method,  
Section~\ref{sec:results} presents our results, and Section~\ref{sec:conclusion} concludes our paper. 

Regarding the notation, $a$, $\mathbf{a}$, $\mathbf{A}$ and $\mathbfcal{A}$ represents, respectively, scalars, column vectors, matrices and $D$-dimensional tensors. The superscripts $^T$, $^H$, $^*$ denote, respectively, transposition, Hermitian transposition, and complex conjugation. For a tensor $\mathbfcal{A} \in \mathbb{C}^{M_1 \times M_2 \times \dots M_D}$, $M_d$ refers to the tensor dimension on the $d^\mathrm{th}$ mode.
A $d$-mode unfolding of a tensor is written as $[\mathbfcal{A}]_{(d)} \in \mathbb{C}^{M_d \times M_{d+1} \dots M_D M_1 \dots M_{d-1}}$ where all $d$-mode vectors are aligned as columns of a matrix. The $d$-mode vectors of $\mathbfcal{A}$ are obtained by 
varying the $d^\mathrm{th}$ index from $1$ to $M_d$ and keeping all other indices fixed.
Moreover, $\mathbfcal{A} \times_d \mathbf{U}$ is the $d$-mode product between a $D$-way tensor $\mathbfcal{A} \in \mathbb{C}^{M_1 \times M_{2} \dots \times M_D}$ 
and a matrix $\mathbf{U} \in \mathbb{C}^{J \times M_d}$. The $d$-mode product is computed by multiplying $\mathbf{U}$ with
all $d$-mode vectors of
$\mathbfcal{A}$. In addition, $\mathbfcal{A} \sqcup_d \mathbfcal{B}$ denotes the concatenation of $\mathbfcal{A}$ and $\mathbfcal{B}$ among the $d^\mathrm{th}$ mode. The concatenation $\sqcup_d$ operation also applies to matrices.

\section{MIMO OFDM Scenario}
\label{sec:scenario}
As our scenario we consider environments with many scatterers, i.e., urban macro and urban micro, in frequency bands below 6~GHz. 
In a MIMO OFDM system, the base station (BS) and the user equipment (UE) are equipped with uniform linear arrays (ULAs) with $M_T$ and $M_R$ antennas, respectively. Moreover, there are $N_\mathrm{sub}$. 
In a fixed time slot, the MIMO channel at each sub-carrier $\mathbf{H}_{n_\mathrm{sub}} \in \mathbb{C}^{M_R \times M_T}$ is modeled as
\begin{equation}
    \label{eq:mimochannel}
\mathbf{H}_{n_\mathrm{sub}} = \sum_{i=1}^{L} \alpha_i e^{-j 2 \pi
\left (f_c + \frac{(n_\mathrm{sub}-1)}{N_\mathrm{sub}} \right) \tau_i} \mathbf{a}_R(\theta_i) \mathbf{a}_T(\phi_i)^H,    
\end{equation} where
$f_c$ is the carrier frequency, $n_\mathrm{sub}$ 
is the sub-carrier index,
$L$ is the number of 
MPCs, $\tau_i, \alpha_i$, $\theta_i$, and $\phi_i$ are, respectively, the delay, complex amplitude, direction of arrival (DoA), and direction of departure (DoD) of the $i^\mathrm{th}$ MPC. The ULA steering vector at the receiver side $\textbf{a}_R$ is modeled as
\begin{equation}
\textbf{a}_R(\theta_i) = [1, e^{j\mu_i}, e^{j2\mu_i}, \ldots e^{j(M_R-1)\mu_i}]^T,   
\label{eq:steer}
\end{equation} where  
$ \mu_i = \frac{2\pi}{\lambda}\Delta_d \cos{\theta_i}$, is the spatial frequency and $\Delta_d=\frac{\lambda}{2}$ is the spacing between the antenna elements. The ULA steering at the transmitter side $\textbf{a}_T$ is modeled in a similar way. 

Moreover, we assume that the transmitter uses a fixed grid of beams (GoB)~\cite{14ObaraGoB,16RakashGoB} as beamformers, and that the transmitter beam has already been selected, i.e., the beam management procedures one and two (P-1, P-2) have been performed~\cite{17Rel143gpp}. Therefore, without loss of generality, our channel model can be simplified as a SIMO OFDM wireless channel $\mathbf{H}~\in~\mathbb{C}^{M \times N_\mathrm{sub}}$ at the receiver side which is equipped with an ULA of $M$ antenna elements.
The channel at each sub-carrier $\mathbf{h}(n_\mathrm{sub})$ is modeled as      
\begin{equation}
\mathbf{h}(n_\mathrm{sub}) = \sum_{i=1}^{L} \alpha_i e^{-j 2 \pi
\left (f_c + \frac{(n_\mathrm{sub}-1)}{N_\mathrm{sub}} \right) \tau_i} \mathbf{a}_R(\theta_i) + \mathbf{z}(n_\mathrm{sub}), 
\label{eq:channel}
\end{equation} 
where $\mathbf{h}(n_\mathrm{sub})$ is a column of $\mathbf{H}~\in~\mathbb{C}^{M \times N_\mathrm{sub}}$
and $\mathbf{z}(n_\mathrm{sub})~\in~\mathbb{C}^{M \times 1}$ is a zero mean circularly symmetric complex Gaussian noise process.

Due to the high correlation between the MPCs, we apply spatial smoothing~\cite{85shanSpatial} in the sub-carriers dimension.
From the $N_\mathrm{sub}$ sub-carriers, we take $K$ of them to compute the smoothing. Therefore, the new sub-carriers dimension is $N_\mathrm{sub}' = N_\mathrm{sub} - K + 1$. 
The selection matrix for the $k^\mathrm{th}$ smoothing sub-block is defined as
\begin{equation}
    \mathbf{J}_k = 
    \begin{bmatrix}
    \mathbf{0}_{(N_\mathrm{sub}',~k-1)} 
    & \mathbf{I}_{N_\mathrm{sub}'} & \mathbf{0}_{(N_\mathrm{sub}',~K-k)} 
    \end{bmatrix} \in \mathbb{R}^{N_\mathrm{sub}' \times N_\mathrm{sub}},
\end{equation}
and the 
smoothed channel 
tensor $\mathbfcal{H}$ is computed by 
\begin{equation}
\label{eq:smooth}
    \mathbfcal{H} = \left[
    \mathbf{H} \mathbf{J}_1^T ~ \sqcup_3 ~ \mathbf{H} \mathbf{J}_2^T ~ 
    \dots ~\sqcup_3 ~ \mathbf{H} \mathbf{J}_K^T 
    \right] \in \mathbb{C}^{M \times N_\mathrm{sub}' \times K},
\end{equation}
where the $K$ channel smoothed matrices are concatenated in the third dimension, such that $\mathbfcal{H} \in \mathbb{C}^{M \times N_\mathrm{sub}' \times K}$ is our 3-dimensional channel tensor. 

\section{ML for Model Order Selection}
\label{sec:method}
Inspired by LaRGE~\cite{21KorobkovLaRGE} and Unitary Tensor ESPRIT~\cite{08HaardtTensor}, we propose a 
ML method designed as a multi-label classification task which observes all the $d$-mode wireless channel singular values and classifies its multidimensional input as signal or noise singular value. Each output neuron has values ranging between $0$ and $1$. The closer to $1$, the more confident the NN is that the neuron represents a signal singular value. Hence, we set a threshold $\xi$ to define the decision region. Therefore, neurons with output above $\xi$ are signal singular values, which are later summed to express the model order. In the following, we present the data pre-processing and the implementation details of our ML architecture and training. 

\subsection{Data pre-processing}
First, we compute the forward-backward averaged version of the channel tensor $\mathbfcal{H}$ as~\cite{08HaardtTensor}
\begin{equation}
    \mathbfcal{Y} \doteq 
    \left [ 
    \mathbfcal{H}\sqcup_3 
    \left(
    \mathbfcal{H}^* \times_1 \mathbf{\Pi}_{M} 
    \times_2 \mathbf{\Pi}_{N_\mathrm{sub}'} \times_3 
    \mathbf{\Pi}_K
    \right)
    \right] 
\end{equation}
where  
$\mathbf{\Pi}_p$ is a $p \times p$ exchange matrix with ones on its anti-diagonal and zeros otherwise, and $\mathbfcal{Y} \in \mathbb{C}^{M \times N_\mathrm{sub}' \times 2K}$. 

Second, 
we take advantage of the centro-Hermitian characteristics of the forward-backward averaged tensor $\mathbfcal{Y}$, as in Unitary Tensor ESPRIT~\cite{08HaardtTensor}, and apply a real data transformation $\mathbfcal{F} = \varphi (\mathbfcal{Y}) \in \mathbb{R}^{M \times N_\mathrm{sub}' \times 2K}$ which is computed as
\begin{equation}
    \varphi (\mathbfcal{Y}) = \mathbfcal{Y} \times_1 \mathbf{Q}_{M}^H \times_2 \mathbf{Q}_{N_\mathrm{sub}'}^H \times_3 \mathbf{Q}_{2K}^H,
\end{equation}
where $\mathbf{Q}_p \in \mathbb{C}^{p \times p}$ is a left-$\mathbf{\Pi}$-real matrix, i.e., $\mathbf{\Pi} \mathbf{Q}_p^* = \mathbf{Q}_p$. In this way, we avoid to compute covariance matrices, and reduce the complexity by working with real numbers only.

Third, we compute the HOSVD
of $\mathbfcal{F}$ as 
\begin{equation}
    \mathbfcal{F} = \mathbfcal{S} \times_1 \mathbf{U}_1
    \times_2 \mathbf{U}_2 \times_3 \mathbf{U}_3,
\end{equation}
where $\mathbfcal{S} \in \mathbb{R}^{M \times N_{sub}'  \times 2K}$ is the core tensor, and 
$\mathbf{U}_1 \in \mathbb{R}^{M \times M}$,
$\mathbf{U}_2 \in \mathbb{R}^{N_{sub}' \times N_{sub}'}$,
$\mathbf{U}_3 \in \mathbb{R}^{2K \times 2 K}$ are the unitary matrices of the $d$-mode singular vectors, in this case $d = 1, 2, 3$. 
The $d$-mode singular values are computed by the singular value decomposition (SVD) of the $d$-mode unfolding of $\mathbfcal{F}$, as in
\begin{equation}
    \left[ \mathbfcal{F} \right]_{(d)} = \mathbf{U}_d ~
    \mathbf{\Sigma}_d ~ \mathbf{V}_d^H
    ,
\end{equation}
where $\mathbf{U}_d \in \mathbb{R}^{M_d \times M_d}$, $\mathbf{V}_d \in \mathbb{R}^{\tilde{M}_d \times \tilde{M}_d}$ are unitary matrices, and $\mathbf{\Sigma}_d \in \mathbb{R}^{M_d \times \tilde{M}_d}$ has the $d$-mode singular values $\sigma_i^{(d)}$ on its main diagonal, and $\tilde{M}_d = \frac{M N_{sub}' 2K}{M_d}$. 

As our channel model is 3-dimensional, we compute the SVD in each of the three unfoldings of $\mathbfcal{F}$, where 
\begin{equation}
\begin{split}
\boldsymbol{\sigma}^{(1)} = \mathrm{diag}(\mathbf{\Sigma}_1) \in \mathbb{R}^{M \times 1}, \\ 
\boldsymbol{\sigma}^{(2)} = \mathrm{diag}(\mathbf{\Sigma}_2) \in \mathbb{R}^{N_{sub}' \times 1}, ~\mathrm{and} \\
\boldsymbol{\sigma}^{(3)} = \mathrm{diag}(\mathbf{\Sigma}_3) \in \mathbb{R}^{2K \times 1}   
\end{split}
\label{eq:largesig}
\end{equation}
are the $d$-mode singular value vectors which serve as input to LaRGE~\cite{21KorobkovLaRGE} for computing the global eigenvalues and estimating the model order.

Inspired by LaRGE,
our ML input consists of  scaling the singular values by the logarithmic function $\boldsymbol{\sigma}_s^{(d)} =  \ln (\boldsymbol{\sigma}^{(d)})$, and reshaping
each $d$-mode vector to size $N$. When reshaping
dimensions, we check the sizes of each $\boldsymbol{\sigma}_s^{(d)}$ vector and decide on a common size for all the 
three 
$d$-mode singular values vectors. 
If the size of one of the $3$ dimensions
is much smaller than the other $2$ dimensions, we choose $N$ to be of the same order of magnitude as the bigger dimensions. Hence, we extend the size of the smallest $d$-mode singular value vector by copying its smallest singular value on the extended vector positions.
However, if all the $d$-mode singular value vectors have sizes of a similar order of magnitude, we select $N=\mathrm{min}\{M, N_{sub}', 2K \}$. Therefore, some $d$-mode singular values are filtered out when $M_d>N$. 
Finally, the input matrix to our ML method is 
\begin{equation}
\mathbf{G} = 
\left[ \boldsymbol{\sigma}_{s}^{(1)}(1:N) ~ \sqcup_2 ~ \boldsymbol{\sigma}_{s}^{(2)}(1:N) 
~ \sqcup_2 ~
\boldsymbol{\sigma}_{s}^{(3)}(1:N)
\right], 
\end{equation} where $\mathbf{G}\in \mathbb{R}^{N \times 3}$ for a $3$-dimensional channel tensor. 
From an implementation perspective, this ML method can be applied in parallel for every transmitter beam candidate. However, if we can observe a channel tensor with more dimensions, e.g., including an antenna array at the transmitter side as in equation~(\ref{eq:mimochannel}) such that $\mathbfcal{H} \in \mathbb{C}^{M_R \times M_T \times N_\mathrm{sub}' \times K}$, we propose to concatenate all the available $d$-mode singular value vectors along the second dimension and input them to our ML architecture. 
Similarly, if we can perform measurements of SIMO channels at different carrier frequencies, as in equation~(\ref{eq:channel}), the $d$-mode singular value vectors of each carrier measurement should be concatenated along the second dimension.
Therefore, our generic ML input is called $\mathbf{G}\in \mathbb{R}^{N \times D}$, where $D$ is the total number of available $d$-mode singular value vectors.
In Section~\ref{sec:results}, we show that increasing the number of observable dimensions enhances the classification accuracy.

\subsection{ML architecture and training}
We design a NN for the model order selection task as a supervised learning problem for multi-label classification. The input to our NN is the matrix $\mathbf{G}\in \mathbb{R}^{N \times D}$ with all the vectors of $d$-mode singular values. The NN architecture is presented in Table~\ref{tab:net-struc} where the 1-dimensional convolutional layers are introduced to efficiently process the $d$-mode singular values, see Section~\ref{ssec:complexity}. 
The output of our NN is a vector with $\mathbf{1}$s and $\mathbf{0}$s of size $1 \times N$. Each label `$1$' denotes a signal singular value, while each `$0$' label represents a noise singular value, e.g., $[1~ 1~ 1~ 0~ 0 ~ 0~ 0 ~0 ]$ for $N=8$ and $3$ signal singular values.  

The activation function of the output layer is the sigmoid function $\gamma$, calculated as
\begin{equation}
    \gamma(x) = \frac{1}{1+e^{-x}},
\end{equation}
where $x$ is the input to the layer activation function. 
For the loss function, used to update the gradient descent algorithm, we select the binary cross-entropy which is computed for each $n^\mathrm{th} \in [1, 2, \dots, N]$ output neuron as
\begin{equation}
    \mathcal{L}_n = - \left( y_n \log (\tilde{y}_n) + (1 - y_n)\log(1-\tilde{y}_n) \right)
\end{equation}
where $y_n$ is the label, and $\tilde{y}_n$ is the predicted score, both for the $n^\mathrm{th}$ output neuron. Finally, the total training loss is calculated by
\begin{equation}
    \mathcal{L} = \sum_{n=1}^{N} \mathcal{L}_n.
\end{equation}

\begin{table}[bt!]
\centering
\caption{Description of the NN for model order selection of multidimensional data.}
\label{tab:net-struc}
\resizebox{0.7\linewidth}{!}{%
\begin{tabular}{|c|c|c|c|}
\hline
 Layer & $N_\mathrm{filter}$ & Filter size & Activation \\ \hline
 Conv1D & $8$ & $3$ & ReLU \\ \hline
 Conv1D & $1$ & $3$ & ReLU \\ \hline
 Dense & $8$ & - & ReLU \\ \hline 
 Dense & $8$ & - & ReLU \\ \hline 
 Dense & $N$ & - & Sigmoid \\ \hline 
 \end{tabular}}
\end{table}

Therefore, 
we design a multi-label classification NN that maps the multidimensional singular values at the input to signal or noise singular values at the output. Moreover, our NN output includes a reliability measure.
As each $n^\mathrm{th}$ output neuron can have a value between $0$ and $1$, the NN is more reliable that the $n^\mathrm{th}$ neuron represents a signal singular value when $\tilde{y}_n$ is closer to $1$. 
Hence, we set a threshold $\xi$ to define the decision region, such that $\tilde{y}_n$ is a signal singular value if $\tilde{y}_n>\xi$. Then, we sum the number of output neurons which are above $\xi$ to get an estimate of the model order. 

\subsection{ML computational complexity}
\label{ssec:complexity}

\begin{table}[tb!]
\centering
\caption{Comparison of NN complexity.}
\label{tab:compare}
\resizebox{\linewidth}{!}{%
\begin{tabular}{|c|c|c|c|c|c|}
\hline
\multicolumn{3}{|c|}{Baseline NN } & \multicolumn{3}{|c|}{Our NN } \\
\hline
 Layer & $N_\mathrm{filter}$ & $\#$ parameters & Layer & $(N_\mathrm{filter}, \mathrm{filter~ size})$ & $\#$ parameters \\ \hline
 Dense & $F_3$ & $F_3(ND +1)$ & Conv1D & $(F_1, Q_1)$ & $F_1 (Q_1 D +1)$  \\ \hline
 Dense & $F_4$ & $F_4(F_3 +1)$ & Conv1D & $(F_2, Q_2)$ & $F_2 (Q_2 F_1 +1)$  \\ \hline
 Dense & $N$ & $N(F_4 +1)$ & Dense & $F_3$ & $F_3 (O_2 F_2 +1)$  \\ \hline 
 - & - & - & Dense & $F_4$ & $F_4 (F_3 +1)$  \\ \hline 
 - & - & - & Dense & $N$ & $N(F_4 +1)$  \\ \hline 
 \end{tabular}}
\end{table}

For the design of our NN architecture, we target a structure which works efficiently with $D$-dimensional data. Hence, we proposed the NN architecture in Table~\ref{tab:net-struc}, which avoids a rapid increase on the number of trainable parameter when the number of observable dimensions $D$ increases.
Our NN architecture has been inspired by  ECNet~\cite{20YangMOS}, where only two dense inner layers are used for the task of source number detection. However, our problem design is slightly different from the ECNet~\cite{20YangMOS}, as we input $\mathbf{G}\in \mathbb{R}^{N \times D}$ and output $\mathbf{g}\in \mathbb{R}^{N \times 1}$. Therefore, in Table~\ref{tab:compare}, we present our NN architecture and the baseline NN
using generic parameters. If we set the number of filters as $F = F_1 = F_3 = F_4$, $F_2 = 1$, and the size of each convolutional filter to $Q = Q_1 = Q_2$, the output of the second convolutional layer $O_2$ becomes $O_2 = N - 2 Q +2$. The final number of parameters for our NN is $F^2 + F(2N + Q(D-1) +5) + N + 1$, while for the baseline NN it is $F^2 + F (N(D + 1) + 2) + N$. Since $Q \ll N$, the complexity of our proposed NN architecture does not increase as fast as in the architecture with dense layers only. Hence, our NN architecture is more suitable for multidimensional data.

\section{Simulations and Results}
\label{sec:results}

For the model order selection task, we simulate $3$-dimensional wireless channels with a varying number of MPCs, 
as modeled in Equation~(\ref{eq:channel}). We consider an OFDM system with channel bandwidth of $20$~MHz, each sub-carrier spaced by $15$~kHz, sampling frequency $f_s = 30.72~\mathrm{MHz}$, and one pilot signal every $12$ sub-carriers, and $N_\mathrm{sub}~=~100$ resource blocks.    
Five channel datasets are generated, with $1$ to $5$ MPCs, each with 3000 channel samples. 
The channel datasets follow a Rayleigh distribution, and the delays and 
DoAs are drawn from a uniform distribution. All the MPCs within a dataset have different delay values 
with a maximum delay of $5 T_s$, where $T_s = 1/f_s$,
and DoA values range between $[0^o, 120^o]$.
The SIMO system is parameterized by $M=8,~\Delta_d=\lambda/2$, and a signal to noise ratio (SNR) of $20$~dB.
In addition, we consider three different values for the carrier frequency, in order to test if the classification accuracy increases when more $d$-mode singular values are available. Hence, we have one dataset with varying MPCs at baseband ($f_c =0$), and a second dataset with varying MPCs with two $3$-dimensional channels at downlink ($f_{c_\mathrm{down}} = 2.6$~GHz) and uplink ($f_{c_\mathrm{up}} = 2.8$~GHz) according to Equation~(\ref{eq:channel}).   

\begin{figure}[p]
    \centering
    \includegraphics[width=0.65\columnwidth]{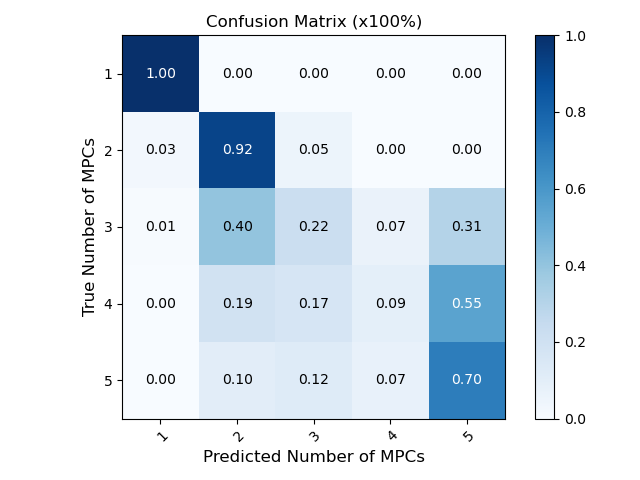}
    \caption{Classification accuracy per class for 1-dimensional ECNet~\cite{20YangMOS}. The input is $\mathbf{g}^{T} \in \mathbb{R}^{50 \times 1}$, where $N=50$ and $f_c = 0$.}
    \label{fig:yang1D}
\end{figure}

\begin{figure}[p]
    \centering
    \includegraphics[width=0.65\columnwidth]{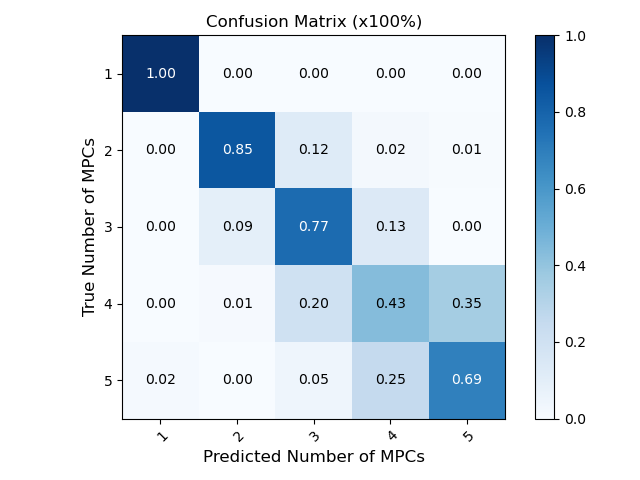}
    \caption{Classification accuracy per class for ECNet~\cite{20YangMOS}. The input is $\mathbf{G} \in \mathbb{R}^{8 \times 6}$, where $N=8$, and the higher-order singular values for $f_{c_\mathrm{down}} = 2.6$~GHz and $f_{c_\mathrm{up}} = 2.8$~GHz are concatenated on the second dimension.}
    \label{fig:yang3D}
\end{figure}

\begin{figure}[p]
    \centering
    \includegraphics[width=0.65\columnwidth]{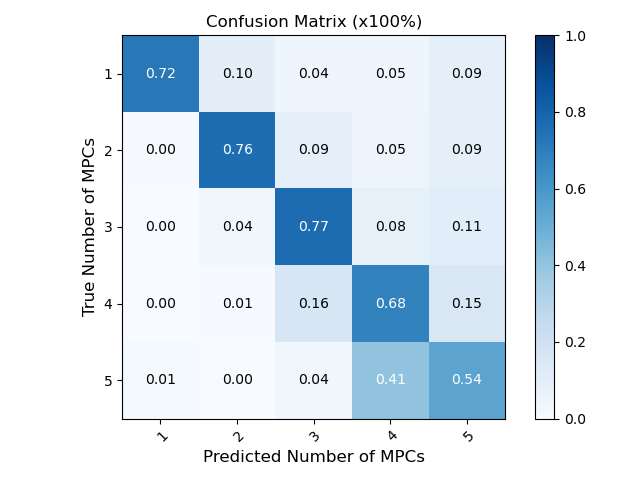}
    \caption{LaRGE~\cite{21KorobkovLaRGE} classification accuracy per class for $\mathbfcal{S} \in \mathbb{R}^{8 \times 51 \times 100}$, the $3$-dimensional channel dataset at $f_c = 0$, and $\rho=0.57$.}
    \label{fig:largebaseband}
\end{figure}

\begin{figure}[p]
    \centering
    \includegraphics[width=0.65\columnwidth]{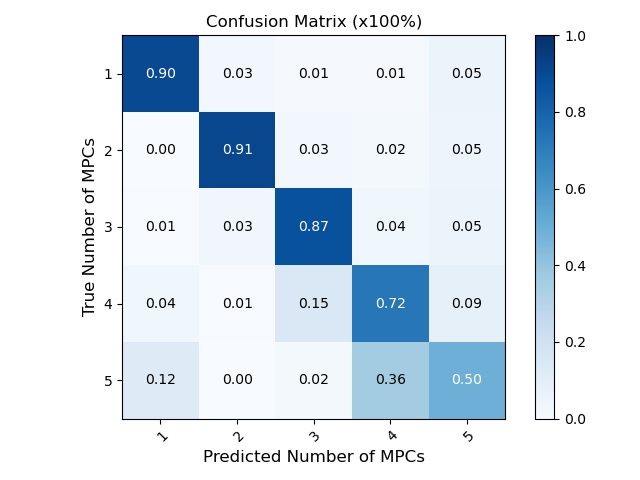}
    \caption{LaRGE~\cite{21KorobkovLaRGE} classification accuracy per class for $\mathbfcal{S} \in \mathbb{R}^{8 \times 51 \times 100 \times 8 \times 51 \times 100}$, the combined channel dataset at $f_{c_\mathrm{down}} = 2.6$~GHz, and $f_{c_\mathrm{up}} = 2.8$~GHz. The decision threshold is $\rho=0.57$.}
    \label{fig:largeupdown}
\end{figure}

\begin{figure}[p]
    \centering
    \includegraphics[width=0.65\columnwidth]{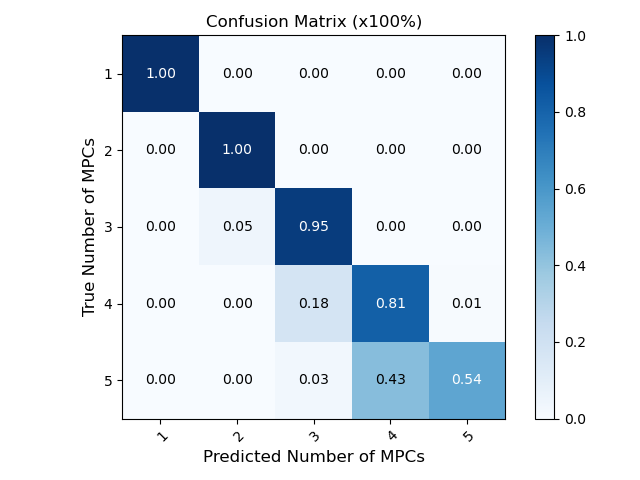}
    \caption{Classification accuracy per class of our proposed ML method. The input is $\mathbf{G} \in \mathbb{R}^{50 \times 3}$, where $N=50$, and the $3$-dimensional channel dataset at $f_c = 0$. The threshold is $\xi=0.8$.}
    \label{fig:ML_3input}
\end{figure}

\begin{figure}[p]
    \centering
    \includegraphics[width=0.65\columnwidth]{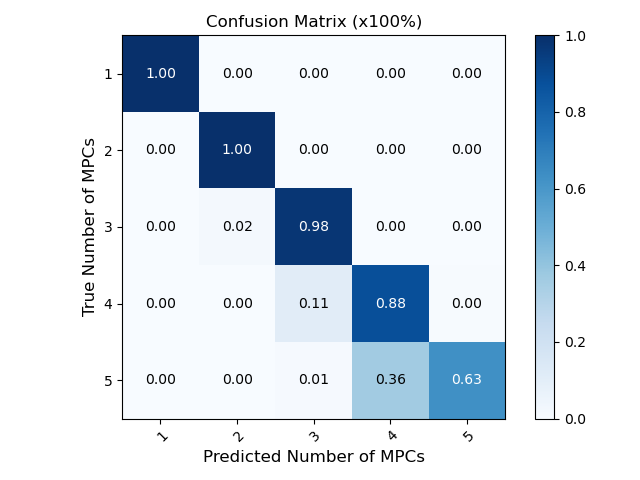}
    \caption{Classification accuracy per class of our proposed ML method. The input is $\mathbf{G} \in \mathbb{R}^{50 \times 6}$, where $N=50$, and concatenated $3$-dimensional channel datasets at $f_{c_\mathrm{down}} = 2.6$~GHz, and $f_{c_\mathrm{up}} = 2.8$~GHz. The threshold is $\xi=0.8$.}
    \label{fig:ML_6input}
\end{figure}

For a performance comparison, we select the ECNet~\cite{20YangMOS} as the ML baseline.
The ECNet was designed to estimate the number of sources using the information of $1$-dimensional eigenvalues. Therefore, we take $\mathbf{h}_b \in \mathbb{C}^{1 \times N_\mathrm{sub}}$, the SISO version of the channel in Equation~(\ref{eq:channel}), and apply smoothing with $K=50$ and $N_\mathrm{sub}' = 51$ on the sub-carriers domain $\mathbf{H}_b \in \mathbb{C}^{N_\mathrm{sub}' \times K}$. 
After that,
we compute the forward-backward averaging.
Then, we compute the singular values in the delay domain $\mathbf{\Sigma}_b \in \mathbb{R}^{N_\mathrm{sub}' \times 2K}$. The input vector to ECNet is $\mathbf{g} \in \mathbb{R}^{50 \times 1}$, with $N=50$. 
Figure~\ref{fig:yang1D} presents the classification accuracy for the ECNet in our $1$-dimensional dataset. It can be observed that estimating the number of sources is especially difficult for $1$-dimensional channels due to the potentially close spacing between the MPCs. 
Figure~\ref{fig:yang3D} shows the improvement on the classification performance of the ECNet when we change its input to the higher order singular values of the tensor channels at uplink and downlink $\mathbf{G} \in \mathbb{R}^{8 \times 6}$, without extension of the $2^\mathrm{nd}$-mode singular values as ECNet did not propose it. Nevertheless, those results are not accurate enough.

As the non-ML baseline, we select LaRGE~\cite{21KorobkovLaRGE} which takes as input the higher order singular values as in Equation~(\ref{eq:largesig}), and the decision threshold is set to $\rho = 0.57$. 
Figure~\ref{fig:largebaseband} presents the results of LaRGE for the $3$-dimensional channels at $f_c=0$, $\mathbfcal{S} \in \mathbb{R}^{8 \times 51 \times 100}$. In addition, Figure~\ref{fig:largeupdown} shows the classification accuracy for LaRGE when coupling 
the higher-order singular values for the channels at downlink and uplink $[ \mathbfcal{S}_\mathrm{up}, ~ \mathbfcal{S}_\mathrm{down} ] \in \mathbb{R}^{8 \times 51 \times 100 \times 8 \times 51 \times 100}$. If compared to the ECNet, LaRGE with the coupled tensor channels as input has a better performance for classifying $2, 3$ and $4$ MPCs.  

Regarding our proposed ML method for model order selection, we implement and train the architecture in Table~\ref{tab:net-struc} using TensorFlow 2.0, Keras and Python. We set $K = 50$, $N_\mathrm{sub}' = 51$, and assert the number of observable higher-order singular values at the NN input to $N = K = 50$.  
The weights of the layers are initialized from a truncated normal distribution with zero mean and standard deviation $\sigma = \sqrt{\frac{1}{m}}$, where $m$ is the input size of each weight layer~\cite{12LecunEfficient}. The initial learning rate is set to $2 \times 10^{-3}$ and the Adam optimizer~\cite{14KingmaAdam} is used. The supervised training runs for 400 epochs with batch size of $128$. For both datasets, at baseband and downlink/uplink, the 15000 dataset samples are divided as $70\%$ for training, and $30\%$ for testing. By our NN design, the NN output is also of size $N = 50$, and the threshold for the signal singular value is set to $\xi = 0.8$.

Figure~\ref{fig:ML_3input} presents the classification accuracy per class for the $3$-dimensional dataset at baseband, $\mathbf{G}^{50 \times 3}$. It can be observed that our proposed ML method is already more accurate than the baselines presented. Moreover, in Figure~\ref{fig:ML_6input} we plot the classification accuracy for the $3$-dimensional dataset at uplink and downlink, $\mathbf{G}^{50 \times 6}$. The results show that increasing the number of higher-order modes which serve as input to our ML also helps to improve its classification accuracy. Moreover, at the threshold of $\xi=0.8$, none of the data samples are classified with more MPCs than they actually have.     

\begin{figure}[tb!]
    \centering
    \includegraphics[width=\columnwidth]{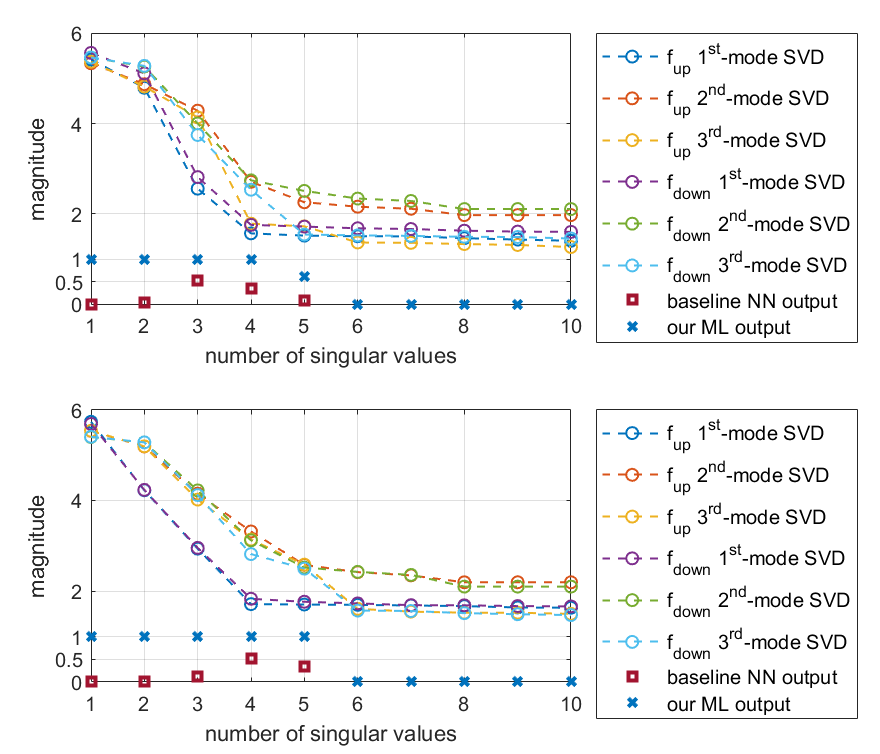}
    \caption{Comparison of two sample classification outcomes for the $5$~MPCs dataset at downlink and uplink. We plot the higher-order singular values which serve as input to the ECNet~\cite{20YangMOS} as well as to our ML method, and the classification output of both ML approaches.}
    \label{fig:mlsample}
\end{figure}

In addition to the higher-order singular values as input, we attribute the success of our proposed ML 
to the design decision of formulating model order selection as a multi-label classification problem. Since each output neuron is classified independently from its neighbors, our ML method could achieve a high reliability. On the contrary, in~\cite{20YangMOS} the classification is modeled as a multi-class classification problem which is easily confused when the dataset has closely spaced MPCs.
Figure~\ref{fig:mlsample} supports this claim, where we plot two sample results for classifying $3$-dimensional channels with $5$~MPCs with downlink and uplink information. The input to both NNs is plotted, as well as their respective classification output. In the channel sample on the top, our ML method classifies it as $4$~MPCs with a high reliability, and the classification would be correct if  $\xi=0.5$. Nevertheless, for the same channel, the ECNet has higher probability for $3$~MPCs, but the probability difference between $3$ and $4$~MPCs classes is small. 
In the channel sample on the bottom of Figure~\ref{fig:mlsample}, our ML method is confident on classifying the input as $5$~MPCs. However, the ECNet is unsure between $4$ and $5$~MPCs and, finally, miss-classify it as $4$~MPCs.
Hence, the ECNet classification performance is not reliable.

\section{Conclusion}
\label{sec:conclusion}
In this paper we propose a ML method for model order selection in MIMO OFDM systems with almost coherent MPCs. 
The results 
shows that the use of higher-order singular values as input to our ML method is effective on improving the classification performance. 
Moreover, our ML design is successful in enhancing the classification accuracy without rapidly increasing the NN computational complexity for multidimensional inputs. 
For future work, we may consider the effect of varying SNRs and to have a ML architecture for model order selection working directly on the channel matrix. 

\section*{Acknowledgement}
This research was partly funded by German Ministry of Education and Research (BMBF) under grant 16KIS1184 (FunKI).

\bibliographystyle{IEEEtran}
\bibliography{mybibfile}

\end{document}